# Exchange interactions and $T_c$ in rhenium doped silicon: DFT, DFT+U and Monte Carlo calculations

Małgorzata Wierzbowska

*Institute of Theoretical Physics, Faculty of Physics,*
*University of Warsaw, ul. Hoża 69, 00-681 Warszawa, Poland*

Interactions between rhenium impurities in silicon are investigated by means of the density functional theory (DFT) and the DFT+U scheme. All couplings between impurities are ferromagnetic except the Re-Re dimers which in the DFT method are nonmagnetic, due to formation of the chemical bond supported by substantial relaxation of the geometry. The critical temperature is calculated by means of classical Monte Carlo (MC) simulations with the Heisenberg hamiltonian. The uniform ferromagnetic phase is obtained with the DFT exchange interactions at room temperature for the impurities concentration of 7 %. With the DFT+U exchange interactions, the ferromagnetic clusters form above room temperature in MC samples containing only 3 % Re.

## I. INTRODUCTION

Silicon doped with rhenium belongs to the widely studied class of materials called diluted magnetic semiconductors (DMS). In this class, the host materials belong to the group III-V, II-VI, and IV semiconductors or transition metal oxides (DMO). Usually the dopants are the 3rd-row $d$-elements, with Mn, Fe and Cr to be most frequently published.[1–6] Interestingly, the 5th-row elements, such as Re, show different properties, especially stronger hybridization with the host, and also deserve investigation.[7] DMS attract much attention, experimental and theoretical, because of their promising application in spintronics.[8–10] To be used in fabrication of electronic devices, materials must preserve their properties in conditions much above room temperature. To date, a very few reports on accessing room temperature has been published, and remain still controversial because Curie temperature depends a lot on the sample preparation.[11–14]

Amorphous silicon doped with rhenium is ferromagnetic at room temparature, but samples loose this property after a few months.[15]

Basic theoretical information about single rhenium impurities in crystalline silicon, in various crystal positions and charged states, have been recently reported.[7] However, ferromagnetism in disordered materials is a complex phenomenon, where single-site impurity characteristics is as important as interactions between the magnetic centers, and also their clustering.[4, 6, 16–19] Therefore, here, a one step closer to the experiment is done, and the work focus on magnetic interactions between two Re impurities embeded in the silicon substitutional sites. The calculations are performed within the density functional theory.[20] Simultaneously, the effect of Coulomb interactions is investigated by means of the DFT+U scheme,[21] since it is well known that the strong interactions within the $d$-shell favour Hund's rules, change the local magnetizations and the binding between impurities, and they strongly influence also the anisotropy of inter-actions, as well as complex observables like the critical temperature.[6, 22] In the next step, the exchange interactions, obtained within the DFT and the DFT+U methods, are used in Monte Carlo simulations of the Heisenberg model and the Curie temperature is estimated.

The main findings concern the geometric, magnetic and statistical effects. In physics of the short-distance interacting pairs, the local geometry plays a very important role, since two rhenium atoms attract each other and form the chemical bonds, which in turn affect very much magnetizations. Therefore, these cells are relaxed in the calculations. Further, the ferromagnetic (FM) and the antiferromagnetic (AF) energy difference is examined as a function of the Re-Re distance and orientation in the crystal. In both theoretical methods, the FM interactions are lower in the energy. Moreover, in the DFT+U scheme the magnetic couplings are stronger and very anisotropic. The mechanism of magnetic interactions between Re impurities in silicon is the ferromagnetic superexchange. The formation energies calculated for all pairs lead to the close pair scenario with a very few single impurities. Clusters with more than two rhenium atoms are out of scopes of this theoretical studies, but in the future the agregates of the substitutional and the interstitial impurities will be investigated too. Finally, the transition from a magnetically disordered to an ordered phase has been studied by means of MC simulations. It was found that, this transition was different when the magnetic interactions calculated with the DFT or the DFT+U methods were embeded to the model. Two main results are obtained: i) the uniform growth of the total magnetization with decreasing temperature using the pair interactions calculated within the DFT method, and ii) formation of magnetic clusters, when the DFT+U interactions were implanted into the MC simulations, similarly to the effect of superparamagnetic blocking.[6]

This paper is organized as follows: in section II the computational details are given, in section III the results for close- and medium-distance pairs are collected, in section IV the long range effects and magnetic anisotropy are described together with the formation energies of all Re-Re pairs, in section V the implication of calculated exchange interactions for the Curie temperature is discussed, the summary is given in section VI.

## II. COMPUTATIONAL DETAILS

The *ab initio* calculations of total energies were performed within the density functional theory,[20] using the plane-wave pseudopotential code QUANTUM ESPRESSO.[23] The scalar relativistic pseudopotential for rhenium has been generated for the valence configuration $6s2\ 5d5$, whithin the ultrasof pseudopotential (USPP) scheme,[24] including the nonlinear core correction.[25] The DFT functional, within the generalized gradient approximation (GGA) of Perdew-Wang (PW91) type,[26] has been employed for calculations of the electronic structure of rhenium pairs in many large supercells. All calculations were repeated with the GGA+U method,[21] accounting for the strong correlations effects. The effective U parameter (in fact U-J) has been estimated from the response method,[27] and its value of 2.3 eV is the same as in the previous work.[7]

The plane-wave cutoffs for the energy of 30 Ry and for the density of 300 Ry are sufficient for the USPP without the semicore electrons in the valence calculations. For the smearing at the Fermi surface, the metallic broadening[28] of 0.01 Ry is used. The k-point mesh $4 \times 4 \times 4$ has been checked, to give the same results as the twice denser mesh.

All supercells were constructed by repetition of the simple cubic 8-atom cell. The supercell sizes were: $2 \times 2 \times 2$ (64 atoms, the concentration of Re was $x$=3.125 %), $3 \times 2 \times 2$ (96 atoms, $x$=2.083 %), $4 \times 2 \times 2$ (128 atoms, $x$=1.5625 %), $3 \times 3 \times 2$ (144 atoms, $x$=1.3888 %) and $4 \times 4 \times 2$ (256 atoms, $x$=0.78125 %). Above procedure has been applied in order to gather all possible directions with a minimal computational cost. The first Re atom was placed at the crystal site $(0, 0, 0)$, and the second impurity was at the site $(k, l, m)$ in units of the silicon lattice constant divided by four. We did not account for interactions of the impurities implanted in the calculated cells with their periodic images (atoms from the neighboring cells), since for neutral cells these effects are not very pronounced.[29, 30] All calculations were performed at the lattice constant of 10.32 a.u., which was optimized with the applied pseudopotentials.

Monte Carlo (MC) simulations of the thermal average of magnetization ($\bar{M}$) were performed using the Metropolis algorithm[31] with the Heisenberg hamiltonian:

$$H = -\frac{1}{2} \sum_{ij} J_{ij} S_i S_j. \tag{1}$$

The interaction constants $J_{ij}$ act between two impurities at sites $i$ and $j$, and they are equal to the total energy difference in the antiferromagnetic and the ferromagnetic state ($E_{AF} - E_{FM}$). Numbers $S_i$ and $S_j$ are magnetic moments (+1 or -1) at the doped sites.

The cells for the simulations have been constructed from $L \times L \times L$ replicas of the 8-atom simple cubic cell. The periodic boundary conditions were applied in MC supercells. The critical temperatures ($T_C$) were estimated from the crossing of the 4th-order Binder cumulant[32] curves, defined as $U_L(T) = 1 - \langle M^4 \rangle / (3 \langle M^2 \rangle^2)$, and calculated in cells of the size $L$ =16, 20 and 24. The number of MC measurements was 5000 sweeps per impurity, and for the warming 500 sweeps per impurity. Some of MC simulations were repeated with higher number of sweeps per impurity, equal to 50000, for both the warming and the MC measurements, to check the convergence. The sites for the impurities have been chosen randomly in 3D supercells according to an algorithm of the 3rd-order Sobolev's chains,[33] in order to obtain a quasi uniform random distribution. At the end, the average over 32 random geometries was taken, to find the magnetization and the $U_L$ curves. Simulations have been performed at the concentrations of 3, 5 and 7 % for the GGA pair interactions, and at lower concentrations of 2, 3 and 5 % for the GGA+U interactions (in this method the rhenium ions couple much stronger than in the GGA).

## III. RESULTS FOR CLOSE- AND MEDIUM-DISTANCE PAIRS

In our previous work,[7] the lattice relaxation around the substitutional site was negligible, but for the impurities which interact on a short distance the geometry optimization is important. Therefore, the positions of atoms in the cells with pairs: 111, 220, 11$\bar{3}$, 331 and 33$\bar{3}$ were optimized within the Newton-Raphson optimization scheme based on the Broyden-Fletcher-Goldfarb-Shanno (BFGS) algorithm.[34] It is found that the optimized Re-Re distances for the close pairs are shorter, and that for medium separations these distances almost do not change; the corresponding numbers are: 6.4 % for the 111 pair, 23.6 % for 220, 3 % for 11$\bar{3}$, 0.2 % for 331 and 0.05 % for 33$\bar{3}$. From now, the relaxed pairs are denoted by "R", and results for the cells with considerable relaxations, e.g. 111R, 220R and 11$\bar{3}$R, will will be presented in detail.

Table I collects the total and the absolute magnetizations per Re and impurities separations in the cells with short- and medium-distance pairs. In the GGA method, the very close pairs, 111 and 111R, are nonmagnetic. The pair 220R also converges to the nonmagnetic state, if one starts from the AF state, while in the FM state it is magnetic. The nonrelaxed pair 220 is magnetic in both the FM and the AF states, however the magnetization of the AF state is largely reduced. The same situation with even stronger reduction of local magnetic moment is in the relaxed pair 11$\bar{3}$R. In contrast, within the GGA+U method, almost all pairs are magnetic in both the FM and the AF state, except the pairs 111 and 111R which are nonmagnetic when one starts from the AF state. All magnetizations in the GGA+U method are higher than in the GGA method. This is usual behaviour, due to the larger exchange splitting withhin the $e_g$ states of rhenium; see the density of states (DOS) of a single substitutional Re in the previous calculations.[7]

| pair | $R_{pair}$ | $M_{FM}$ | | $M_{AF}$ | |
|------|-----------|----------|------|----------|------|
| | | GGA | GGA+U | GGA | GGA+U |
| 111R | 4.18 | 0.00/0.00 | 1.57/1.76 | 0.00 | 0.00 |
| 111 | 4.47 | 0.00/0.00 | 1.83/2.12 | 0.00 | 0.00 |
| 220R | 5.57 | 1.29/1.45 | 1.96/2.47 | 0.00 | 2.90 |
| 220 | 7.30 | 1.86/2.06 | 1.99/2.49 | 1.14 | 2.04 |
| 113R | 8.30 | 1.10/1.20 | 1.99/2.30 | 0.85 | 1.69 |
| 11$\bar{3}$ | 8.56 | 1.70/1.84 | 1.99/2.29 | 1.42 | 1.86 |
| 400 | 10.32 | 1.86/2.04 | 2.00/2.38 | 1.44 | 2.12 |
| 331 | 11.25 | 1.76/1.89 | 1.99/2.33 | 1.40 | 2.10 |
| 33$\bar{3}$ | 13.41 | 1.81/1.98 | 1.98/2.32 | 1.82 | 2.24 |

TABLE I. Total/absolute magnetizations (in $\mu_B$) in the cells in the FM state, and absolute magnetizations for the AF state, and the impurities separations $R_{pair}$ (in a.u.)

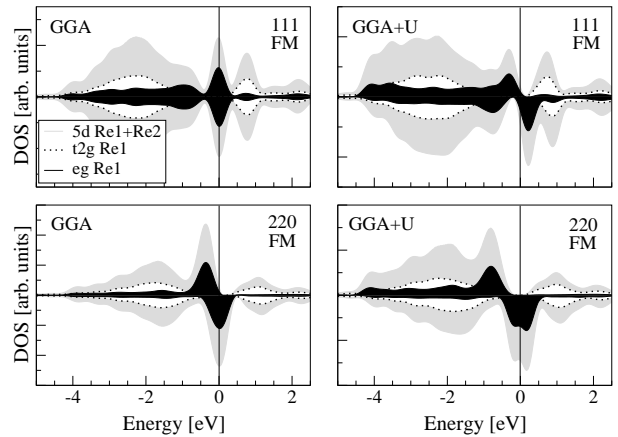

FIG. 1. The density of states (DOS) for the 111 and 220 pairs in the FM/nonmagnetic states obtained within the GGA and the GGA+U methods. The $t_{2g}$ and $e_g$ states for one of the impurities in the pair are drawn on the gray color background of the 5d-states of both Re atoms. The Fermi level is marked by the vertical line.

In this point, it is worth to comment on nonmagnetic states in the close pairs: the substitutional Re in silicon occupies halfly the minority spin $e_g$ state, and this is different from the homovalent element Mn, where the crystal field splitting is weaker than the exchange splitting and the $t_{2g}$ states are at the Fermi level.[7] In the Mn doped silicon, the close pairs are more strongly magnetized than the separate pairs, due to partial occupation of majority spin bands. When the minority spin bands are partially occupied, then the close interactions increase filling of these bands to more "closed-shell" scenario, and the spin polarization decreases.

The density of states for the nonmagnetic solution of the ferromagnetic start for the 111 pair and for the FM phase of the 220 pair obtained within the GGA and GGA+U methods are presented in Figure1. The strong hybridization of 5d-Re states with the neighboring atoms is obvious for the 111 pair. In the case of 220 pair, the degeneracy of the $e_g$ spin-down states is partially lifted due to the broken symmetry around the Re atom when the second impurity is placed very closely.

## IV. LONG-DISTANCE PAIRS AND MAGNETIC ANISOTROPY AND FORMATION ENERGIES

This section is focused on general trends in magnetic interactions for all pairs, and on the formation energies with respect to the most bound pair 111R; Table II gathers all these numbers.

The formation energy of all pairs (except 111, which however is only given for the comparison, to illustrate the relaxation effect) cost about 1 eV more than formation of the close 111R dimer. The fact, that the GGA+U formation energies are lower than those from the GGA, does not change the scenario of very close pairs. At this point, one could say, that it is not worth to deal with the rest of pairs because statistically there will be a very few of them in the sample. It would be reasonable to focus on clusters with more than two impurities and it will be a subject of the future work. Now, the conclusions from medium- and long-distance pairs will be drawn, since in the large sample they also exist and mediate magnetic interactions for a long distance, as will be discussed in connection with the critical temperature.

The magnetic couplings $J_{int}=E_{AF}-E_{FM}$, obtained with the GGA method, decay quickly with the distance, and only the pairs in the 64 atom cell are strongly coupled, with the maximal exchange interaction found for the 220 pair. All pairs are coupled ferromagnetically (positive numbers $J_{int}$) in both applied here methods.

As for the anisotropy of magnetic interactions, the GGA numbers do not show any visible dependence on the crystal direction. While among the GGA+U values, there is well pronounced enhancement of the coupling along the 110 crystal axis. This direction is prefered for the couplings in zinc-blende hosts, due to the zygzak chains, which mediate magnetic interactions via a small polarization of the host atoms neighboring to impurities. These polarizations are visible in the Löwdin population analysis,[35] and amount to about 0.01÷0.02 of electron charge per the Si site.

The range of magnetic interactions in the GGA+U method is long. There are quite large $J_{int}$ numbers along 110-direction for the distance over 20 a.u.; these are for instance numbers 23 meV for the 660 pair, and about 6 meV for the 880 pair in the 256-atoms cell. At long-distances, even small couplings play a very crucial role in estimation of the Curie temperature, because they permit a percolation between impurities at small concentrations.[36] For instance, it is known, that to get FM-order at concentration of 7 %, the exchange interactions must extend to at least five coordination shells.[6]

| pair | $N_{pair}$ | $R_{pair}$ | $J_{int}$ | | $E_f$ | |
|------|-----------|-----------|-----------|---------|-------|---------|
| | | | GGA | GGA+U | GGA | GGA+U |
| **64 atoms cell** | | | | | | |
| 111R | 4 | 4.18 | 0.00 | 23.69 | – | – |
| 111 | 4 | 4.47 | 0.00 | 65.30 | 224 | 143 |
| 220R | 12 | 5.57 | 8.43 | 235.87 | 1171 | 811 |
| 220 | 12 | 7.30 | 40.41 | 139.80 | 1325 | 965 |
| 113R | 12 | 8.30 | 2.18 | 96.31 | 1290 | 971 |
| 11$\bar{3}$ | 12 | 8.56 | 7.60 | 84.42 | 1340 | 998 |
| 331 | 12 | 11.25 | 9.68 | 46.24 | 1267 | 969 |
| 400 | 6 | 10.32 | 19.28 | 62.19 | 1220 | 877 |
| 422 | 24 | 12.64 | 10.31 | 40.32 | 1268 | 967 |
| 33$\bar{3}$ | 4 | 13.41 | 4.22 | 10.66 | 1284 | 985 |
| 440 | 12 | 14.60 | 4.32 | 46.82 | 1237 | 946 |
| 444 | 8 | 17.88 | 2.62 | 4.52 | 1296 | 996 |
| **96 atoms cell** | | | | | | |
| 511 | 12 | 13.41 | 5.40 | 14.07 | 1233 | 930 |
| 53$\bar{1}$ | 24 | 15.26 | 7.43 | 21.94 | 1234 | 931 |
| 620 | 24 | 16.32 | 6.57 | 17.82 | 1230 | 927 |
| 533 | 12 | 16.92 | 2.96 | 7.72 | 1236 | 937 |
| 642 | 48 | 19.31 | 3.60 | 8.47 | 1240 | 941 |
| **128 atoms cell** | | | | | | |
| 71$\bar{1}$ | 12 | 18.24 | 3.56 | 10.52 | 1200 | 903 |
| 731 | 24 | 19.82 | 2.06 | 4.54 | 1203 | 904 |
| 800 | 6 | 20.64 | 1.75 | 11.04 | 1201 | 906 |
| 73$\bar{3}$ | 12 | 21.12 | 1.22 | 3.06 | 1204 | 906 |
| 822 | 24 | 21.89 | 1.67 | 3.69 | 1203 | 905 |
| 840 | 24 | 23.08 | 1.33 | 2.85 | 1200 | 901 |
| 844 | 24 | 25.28 | 0.69 | 0.89 | 1204 | 907 |
| **144 atoms cell** | | | | | | |
| 551 | 12 | 18.43 | 6.68 | 19.31 | 1182 | 886 |
| 55$\bar{3}$ | 12 | 18.78 | 1.13 | 3.03 | 1192 | 898 |
| 555 | 4 | 22.34 | 1.13 | 3.05 | 1192 | 898 |
| 660 | 12 | 21.89 | 6.74 | 23.31 | 1182 | 887 |
| 664 | 24 | 24.20 | 0.64 | 0.70 | 1191 | 898 |
| **256 atoms cell** | | | | | | |
| 75$\bar{1}$ | 24 | 22.34 | 0.43 | 1.07 | 1107 | 818 |
| 753 | 24 | 23.50 | 1.04 | 2.19 | 1105 | 816 |
| 771 | 12 | 25.67 | 1.93 | 4.34 | 1109 | 813 |
| 77$\bar{3}$ | 12 | 25.93 | 0.30 | 0.64 | 1103 | 813 |
| 880 | 12 | 29.19 | 0.65 | 5.60 | 1098 | 812 |

TABLE II. Interactions $J_{int}$ (in meV) between rhenium pairs. We assumed $J_{int}$ being negligible for pairs: 75$\bar{5}$, 775, 777, 862, 866, 884, 888 (256 atoms cell). $E_f$ is the formation energy (in meV) of the lower-energy magnetic phase for the given pair, with respect to the formation energy of the dimer 111R. The distance $R_{pair}$ (in a.u.) between Re-Re, and the coordination number $N_{pair}$ of given distance impurities within the shell around the impurity ion at (0,0,0) are also given.

| $x$ | T$_C^{MC(\bar{M})}$ | T$_C^{MC(U_L)}$ | T$_{M>0}^{MC}$ |
|-----|--------------------|-----------------|----------------|
| 3% | 84 | 86 | 69÷116 |
| 5% | 190 | 197 | 185÷220 |
| 7% | 290 | 301 | 278÷330 |

TABLE III. Curie temperatures (in K) at various concentrations $x$ obtained from the MC simulations by: 1) the magnetization curves crossing T$_C^{MC(\bar{M})}$ and 2) the cumulant curves crossing T$_C^{MC(U_L)}$, of averaged samples, using the GGA interactions $J_{int}$ from Table II. T$_{M>0}^{MC}$ is the temperature range at which the magnetization in studied samples starts to rise from zero.

## V. CRITICAL TEMPERATURES

In this section, the discussion of the results of Monte Carlo simulations with the Heisenberg model is presented. The simulations were performed on a number of supercells with different sizes, and with various impurities distributions, and for a few concentrations. In the hamiltonian, the Re-Re interactions $J_{int}$, calculated within the GGA and the GGA+U schemes (see Table II) have been used.

For the interactions $J_{int}$ obtained within the GGA method, which are rather not anisotropic in space, all samples start to magnetize at some temperature, and this magnetization saturates to high values of 0.7÷0.95 per impurity, depending on a sample. The starting temperature for a visible rise of magnetization depends quite strongly on a geometry of impurities distribution. The Curie temperatures for the ferromagnetic order were estimated in two ways: (1) from the crossing of the magnetization curves, and (2) from the crossing of the 4th-order cumulant curves.[32] These curves have been calculated for different supercell sizes, and each of these curves has been averaged over geometries of impurities distribution – as it is described in section with the computational details.

The Curie temperatures, calculated with the aforementioned procedure, are presented in Table III. The temperatures obtained from the magnetization-curves crossing are a bit lower than T$_C$ obtained from the cumulant-curves crossing. The last column shows the temperature at which the magnetization in samples starts to rise from zero (T$_{M>0}^{MC}$), and this temperature is not averaged over the impurities distributions, but it is given as a range to which fall the results from all used geometric samples. This quantity shows how much the theoretical results for given concentration differ just due to different impurities distribution. It gives also an argument in the explanation why the experimental T$_C$, published for the same material and the concentration, vary a lot depending on samples preparation.

In contrast to simulations with the GGA interactions, the GGA+U interactions implanted to the Heisenberg hamiltonian usually do not lead to the total and uniform magnetization in the supercell. Independently on geometric parameters used in these simulations, it is found negligible or very low magnetization of samples (0.1÷0.3 $\mu_B$ per impurity) at most simulation temperatures. Interestingly, instead of homogeneous magnetization within a supercell, the ferromagnetic clusters form. In some samples, there was a transition temperature at which it was found a large total magnetization and below this temperature the magnetic clusters formed. In other samples, there is a smooth transition between magnetically disordered phase and the phase with cluster structure. The magnetization of local clusters is also not full, but grows with the decreasing temperature. One can only approximate the transition temperature by watching the pictures for different samples.

| $x$ | $T_{FM}^{MC}$ |
|-----|---------------|
| 2%  | 110÷180       |
| 3%  | 230÷350       |
| 5%  | 690÷820       |

TABLE IV. Approximate temperature ranges $T_{FM}^{MC}$ (in K) at which the ferromagnetic cluster order starts to form in the MC supercells containing impurities at concentration $x$, and using the interactions $J_{int}$ calculated by means of the GGA+U method (from Table II).

Figure 2 presents magnetization of one chosen geometric sample, calculated using the interactions from the GGA+U method, with impurities concentration of 5 %, at five temperatures: 116 K, 464 K, 696 K, 812 K and 928 K. The sample size is L×L×L×8 with L=20, and it is divided into a few boxes, and each of these boxes is of the size 4×4×4×8; thus, the whole supecell is divided into 5×5×5 pieces. Whithin the each small box, the magnetization over impurities is averaged, and it is done over the spins obtained in the last MC measurement sweep. The 3D supercell is presented in the slices, which cut parallel to the XY-plane, and these cut slices are placed in Figure 2 in columns. The averaged magnetization per impurity can take the value from -1 to 1, therefore the absolute value of box-magnetization is marked by a circle with radius proportional to magnetization, and spin direction (up and down) is depicted by the white and black color.

One can see in Figure 2 that the FM-order transition, in presented sample, occurs between 820 and 900 K. In some other samples (not presented in Figure 2), at the same concentration of 5 %, the estimation of $T_{FM}^{MC}$ is not so easy because one cannot find a temperature at which the total sample magnetize; only the magnetic clusters slowly grow.

It is important to mention, that this scenario with magnetic clusters does not change with a very large number of measurement sweeps, of 50 000 per impurity, and with the same number of warming sweeps in the MC simulations.

In Table IV, there are listed the rouly estimated temperatures at which the cluster FM-order starts to form, $T_{FM}^{MC}$, for concentrations: 2, 3, and 5 %. The range of these temperatures is wide, because the magnetic transition depends on impurities distribution in the sample, like it was the case with the GGA critical temperatures. At the concentrations above 4-5 %, all samples form the magnetic cluster structure above room temperature.

It is known about the percolation that, if impurities distribution is of cluster type and clusters are large, then a superparamagnetic blocking phenomenon becomes important.[6, 37] In the simulations performed in this work, the distributions are uniform netherveless the superparamagnetic blocking occurs, and the reason for this is the magnetic coupling of pairs which acts at a very long range. When during the MC simulations the magnetic clusters form, their internal coupling is so strong, that it costs much energy to flip the whole cluster. Finally, the magnetic relaxation time is very long. Thus, the total magnetization, in some of the samples with the cluster structure, is close to zero over all temperatures, even after 50 000 MC measurements per impurity.

## VI. SUMMARY

In this work, the rhenium pairs embeded in silicon in substitutional sites were calculated by means of the GGA and the GGA+U methods. The formation energy of the closest pair is roughly 1 eV lower than those energies for the rest of double combinations. This supports the cluster scenario with a very few single impurities and medium-distance pairs. It is a common behaviour of impurities in various DMS.[16–19]

The calculated magnetic couplings $J_{int}=E_{AF}-E_{FM}$ are positive for all pairs, and very long range. The GGA couplings are not anisotropic with respect to the crystal axes, in contrast the GGA+U interactions are much stronger along the 110 direction. At small distances, the Re-Re pairs become nonmagnetic for both the AF and FM phases, or only for the AF phase. At higher concentrations this effect is even stronger, but less pronounced in the GGA+U method than in the GGA scheme.

The mechanism which drives this DMS magnetic is the ferromagnetic superexchange, with high critical temperatures excessing room temperature, even for low concentrations of impurities of 7 % for the GGA and 3 % for the GGA+U. Magnetizations of Monte Carlo samples below critical temperatures are uniform for the GGA, and different with magnetic clusters structure for the GGA+U scheme. Quite large and very extended magnetic couplings support percolation in this DMS, and give hope to obtain still high Curie temperature when the impurities clusters are included in simulations.

There is also plaussible the existence of superparamagnetic blocking phenomenon below the Curie temperature. But this scenario must be confirmed by further studies with clusters of the substitutionals and the inter-stitials together, and also employing a method for better treatment of the self-interaction effect in the host (silicon), namely the DFT with the pseudopotential self-interaction correction (pSIC).[38]

## VII. ACKNOWLEDGMENTS

I thank Prof. R. R. Gałązka for showing me the experimental results and to Andrzej Fleszar for many philosophical discussions. The calculations have been performed at the Leibniz Supercomputing Centre in Munich. The work was supported by the European Founds for Regional Development within the SICMAT Project (Contract No. UDA-POIG.01.03.01-14-155/09).

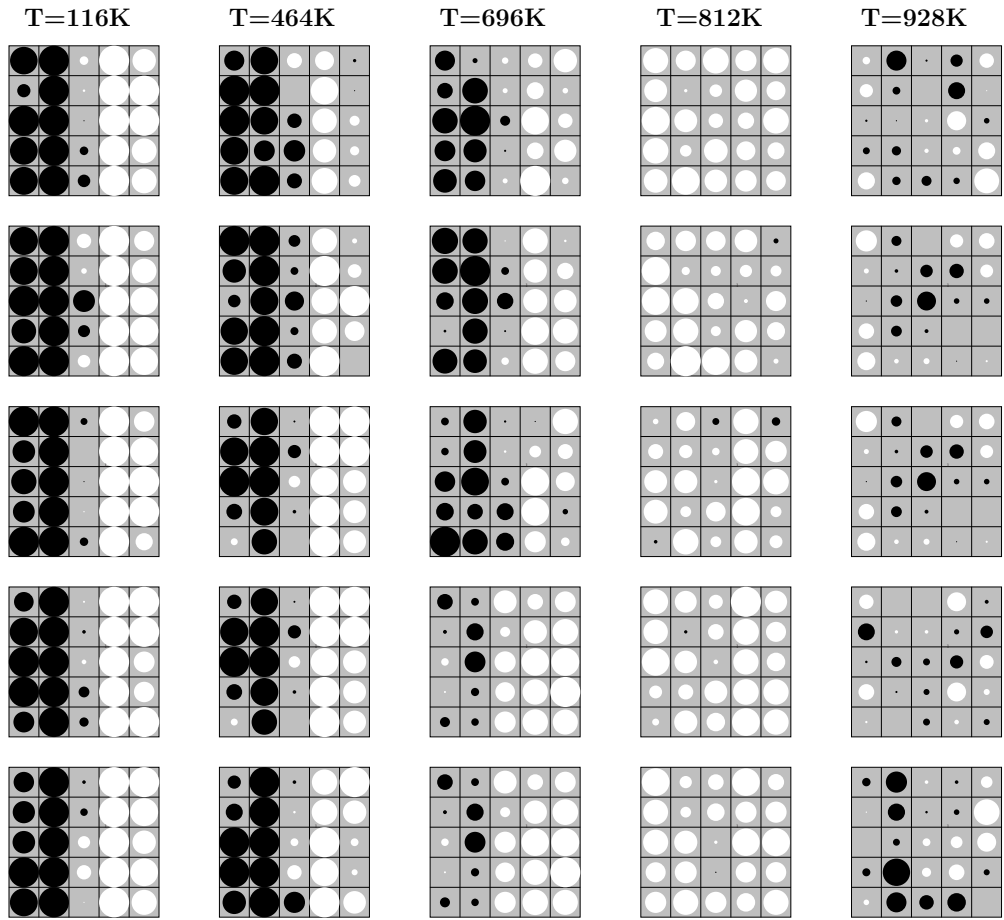

FIG. 2. Ferromagnetic clusters in one chosen MC sample, which contains 5 % Re. Simulations were performed with the GGA+U interactions from Table II and at five temperatures: 116 K, 464 K, 696 K, 812 K, 928 K. Each column presents five cuts through the sample, which are perpendicular to the xy-